\documentclass{article}

\usepackage{spconf,amsmath,graphicx}
\usepackage{booktabs}
\usepackage{multirow}
\usepackage{subfigure}
\usepackage{enumitem}

\title{Speech Dereverberation with A Reverberation Time Shortening Target}
\name{Rui Zhou$^1$, Wenye Zhu$^{1,2}$, Xiaofei Li$^{1,*}$ \thanks{*Corresponding author.}}
\address{
  $^1$Westlake University \& Westlake Institute for Advanced Study, Hangzhou, China\\
  $^2$Zhejiang University, Hangzhou, China}

\begin{document}
\ninept

\maketitle
\begin{abstract}
This work proposes a new learning target based on reverberation time shortening (RTS) for speech dereverberation. The learning target for dereverberation is usually set as the direct-path speech or optionally with some early reflections. This type of target suddenly truncates the reverberation, and thus it may not be suitable for network training. The proposed RTS target suppresses reverberation and meanwhile maintains the exponential decaying property of reverberation, which will ease the network training, and thus reduce signal distortion caused by the prediction error. Moreover, this work experimentally study to adapt our previously proposed FullSubNet speech denoising network to speech dereverberation.  Experiments show that RTS is a more suitable learning target than direct-path speech and early reflections, in terms of better suppressing reverberation and signal distortion. FullSubNet is able to achieve outstanding dereverberation performance. 
\end{abstract}


\section{Introduction}

Severe late reverberation brings significant damage to the quality and intelligibility of speech \cite{Helfer1990Hearing} and will also cause performance degradation for back-end tasks such as automatic speech recognition (ASR) \cite{Yoshioka2012Making}. Normally, early reflections does not cause so much negative effect \cite{Arweiler2011influence}. Speech dereverberation, especially the single-channel case, is still a challenging task.

Before deep neural network (DNN) has been widely used, the traditional dereverberation methods  were based on statistical models and signal processing algorithms. The essential problem of dereverberation is the deconvolution between speech signal and room impulse response (RIR). Deconvolution can be accomplished by applying an inverse filter of RIR to the reverberant speech, which is referred to as inverse filtering methods, e.g. \cite{LiXiaofei2019Multichannel}, \cite{LiXiaofei2018Identification}, \cite{LiXiaofei2018Multisource}. As for the inverse filtering methods, accurate RIR must be first blindly identified, which is very challenging espcially for the single-channel case \cite{Swami1994Multichannel}. Even if the RIR is known, due to its non-minimum phase characteristics in typical cases, directly computing its inverse filter will cause system instability or non-causality \cite{Neely1979Invertibility}, \cite{Miyoshi1988Inverse}. Moreover, inverse filtering is very sensitive to noise. Alternatively, instead of resolving the inverse filter of RIR, weighted prediction error (WPE) \cite{Nakatani2010Speech,Marc2014linear} uses linear prediction to directly estimate the inverse filter from reverberant signal, and applies the inverse filter to remove late reverberation. WPE has achieved remarkable  performance, and is one of the most popular dereverberation methods. Another technical line of dereverberation is spectral subtraction, following the perspective of speech enhancement. Late reverberation can be considered as additive noise, which is assumed to be independent of direct path signal and early reflections \cite{Habets2000Late, Unoki2003Amethod}. In \cite{braun2018evaluation}, methods for estimating the power spectrum density of late reverberation have been summarized.

The application of DNN has made a great progress in solving speech dereverberation. The basic idea is to construct  a nonlinear mapping function, based on supervised learning of DNN, from the spectral feature of reverberant speech to the one of target speech. The input feature could be directly the time-domain signal, or the STFT (short-time Fourier transform) coefficients or magnitude spectrum of reverberant speech. Correspondingly, the output features could be the time-domain signal,  STFT coefficients, magnitude spectrum or magnitude mask of target speech.   
The network architecture used for single-channel speech dereverberation has been evolved a lot, and made a great progress, from the initial fully connected networks \cite{Han2014Learning, Han2015Learning} to recurrent neural networks (RNN) with long short-term memory (LSTM) for time series modeling \cite{Zhao2018Late, Santos2018speech}, and to convolutional neural networks (CNN), such as U-NET \cite{Chun2021Comparison, Wang2020deep} and temporal convolutional networks (TCN) \cite{Zhao2020Monaural}, then to (self-)attention-based methods \cite{Zhao2020NoisyReverberantSE, Helin2021TeCANet}. 

In this work, we experimentally study to adapt our two previously proposed speech denoising networks, i.e. subband LSTM network \cite{li2020online} (refered to as SubNet) and FullSubNet \cite{hao2020fullsubnet}, for speech dereverberation. Based on the cross-band filters model \cite{avargel2007system}, the time-domain convolution between source speech and RIR can be decomposed into subband convolutions, and thence speech dereverberation can be perfectly performed in subband based on deconvolution or inverse filtering. SubNet inputs the noisy spectra of one frequency and its neighbouring frequencies, and outputs/predicts the clean speech spectra of this frequency, which seems exactly suitable for speech dereverberation by mimicking the inverse filtering process. FullSubNet combines SubNet with a fullband network to also exploit the fullband spectral pattern, as the enhanced speech should have a correct spectral pattern across all frequencies. FullSubNet used for speech dereverberation can be seen as a combination of speech spectral regression and subband inverse fitering. Experiments show that SubNet and FullSubNet are indeed able to achieve outstanding dereverberation performance.


More importantly, this work proposes a new learning target based on reverberation time shortening (RTS). DNN-based methods in the literature normally takes the direct-path speech as learning target, which actually is a very strict target as removing all reverberation. As a result, they normally have a large prediction error, which may cause speech distortion. Since early reflections do not cause speech quality degradation, they are often preserved and only late reverberation are removed, such as in WPE \cite{Nakatani2010Speech,Marc2014linear} and the spectral subtraction methods \cite{braun2018evaluation}. Preserving early reflections in the learning target would reduce the prediction error of the network. However, preserving only early reflections will also reduce the sound naturalness, as sounds appear in real life never have such type of reverberation form. In addition, no matter which training target is used, the direct path or early reflections, the network need to learn a sudden truncation of reverberation, which is not fully suitable for network training and will cause signal distortion. The proposed learning target is a shortened version of the original RIR, and has a small target $T_{60}$, e.g. 0.15 s. Instead of suddenly truncating RIR, this target still maintains the property of exponential decay, which will maintain the sound naturalness and also ease the network training. Experiments show that using the proposed learning target can more effectively suppress reverberation and signal distortion.     

\section{The Proposed Method}

Denote single-channel signals in the time domain as:
\begin{equation}
  y(n)=s(n) * a(n) + e(n),
  \label{eq1}
\end{equation}
where * stands for convolution, \(n\) denotes the discrete time index. \(y(n)\), \(s(n)\), \(a(n)\) and $e(n)$ are reverberant speech, clean speech, RIR and ambient noise, respectively. This work mainly works on dereverberation, but certain amount of ambient noise will also be considered and suppressed. 

We can divide RIR \(a(n)\) into two parts, where $a_d(n)$ and \(a_u(n)=a(n)-a_d(n)\) are the desired and undesired parts, respectively. The reverberant speech can be rewritten as:%
\begin{equation}
s(n) * a(n)=s(n) * a_{d}(n)+s(n) * a_{u}(n)=x(n)+u(n)
  \label{eq2}
\end{equation}
This work aims to recover the desired signal \(x(n)\) from the reverberant and noisy speech \(y(n)\).

Setting the learning target as the direct path speech or with some early reflections amounts to applying a rectangular window $w(n)_{\text{rect}}$ to RIR to obtain the dirsired part of RIR, i.e. $a_{d}=w_{\text{rect}}(n)a(n)$. The rectangular window for direct path and 50 ms of early reflections are shown in Fig.~\ref{fig:windows} (a). The rectangular window suddenly truncates the RIR, as shown in Fig.~\ref{fig:windows} (b) and (c) for the target of direct path and early reflections, respectively. This may make the neural network hard to learn a mapping function between the input and the output, and leads to a large prediction error and signal distortion. 

\subsection{Learning Target: Reverberation Time Shortening}

In this work, we propose a new learning target based on RTS, which is a shortened version of the original RIR, and has a small target $T_{60}$, e.g. 0.15 s. Instead of suddenly truncating the RIR, the RTS target still maintains the property of exponential decay, which will maintain the sound naturalness and also ease the network training. 

Formally, we define the new window function as :
\begin{equation}
  w(n)=\left\{\begin{array}{ll}
1 & \text { for } n \leq \mathrm{N}_{1} \\
10^{-q\left(n-N_{1}\right)} & \text { for } n>{N}_{1}
\end{array}\right.
  \label{eq3}
\end{equation}
where \(N_{1}\) denotes the discrete time index when the direct path ends. The parameter \(q\) controls the decaying rate of the window. The original RIR would be shortened by applying this window.

\begin{figure}[t]
  \centering
  \subfigbottomskip=2pt 
  \subfigcapskip=-5pt 
  \subfigure[ ]{
  \includegraphics[width=0.48\linewidth]{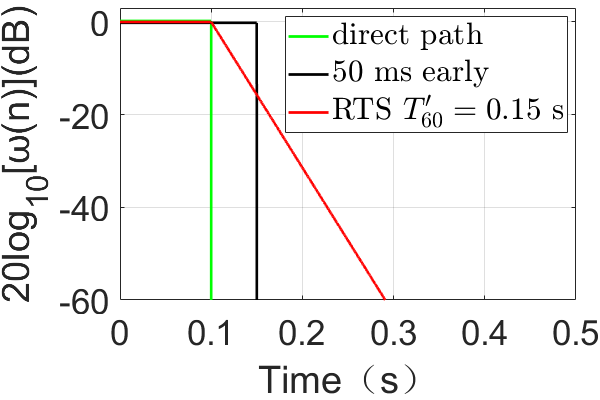}}
  \subfigure[ ]{
  \includegraphics[width=0.48\linewidth]{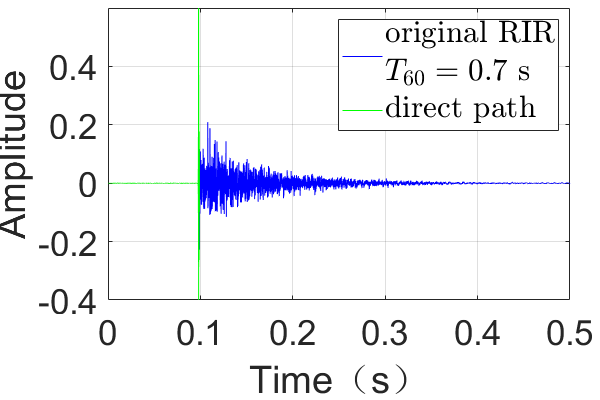}} \\
    \subfigure[ ]{
  \includegraphics[width=0.48\linewidth]{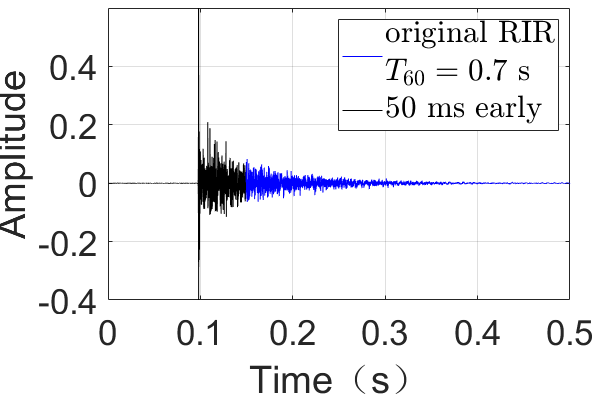}}
  \subfigure[ ]{
  \includegraphics[width=0.48\linewidth]{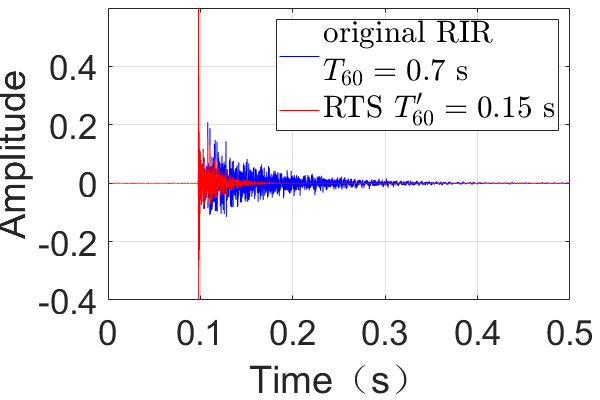}}
  \caption{(a) Window functions. The desired part of RIR for (b) direct path, (c) 50 ms of early reflections, (d) RTS.}
  \label{fig:windows}
  \vspace{-0.0cm}
\end{figure}

In the Polack’s Statistical Model \cite{Naylor2010Speech}, the reverberation component of RIR can be realized by a Gaussian process with an exponentially decaying envelope. Based on this model, RIR can be written in the form of: 
\begin{equation}
a(n)\approx b(n) \mathrm{10}^{-{p} (n-N_1)}  \quad \text { for } n > N_1 
  \label{eq4}
\end{equation}
where \(b(n)\) denotes a zero-mean Gaussian noise sequence, and $p$ reflects the decaying rate. 

Applying the window function to the original RIR, the target 
\begin{align}
a_d(n)&=w(n)a(n) &  \forall n, \nonumber \\ 
&\approx b(n) \mathrm{10}^{-(p+q) (n-N_1)} & \text { for } n > N_1,  
  \label{eq5}
\end{align}
is still exponentially decaying, with a new decaying rate of $p+q$.

Based on the definition of \(T_{60}\), namely power decaying by 60  dB, the original \(T_{60}\) of $a(n)$ and the new \(T_{60}\) of $a_d(n)$ (denoted as \(T'_{60}\)) are respectively
\begin{equation}
{T_{60}} = \frac{3 }{p f_s}, \quad {T'_{60}} = \frac{3 }{(p+q) f_s}, 
  \label{eq6}
\end{equation}
where \(f_{\mathrm{s}}\) denotes the sampling frequency. It is obvious that  \(T'_{60}\) is smaller than \(T_{60}\) as long as $q$ is positive. 

In practice, we set the learning target with a desired \(T'_{60}\). Given the original and target \(T_{60}\)s, the window parameter $q$ is set to: 
\begin{equation}
q=\frac{3}{T_{60}^{\prime}  f_s}-\frac{3}{T_{60}   f_s}.
\label{eq7}
\end{equation}
Fig.~\ref{fig:windows} (a) shows the window function for the case with $T_{60}=0.7$ s and $T'_{60}=0.15$ s, and Fig.~\ref{fig:windows} (d) shows the corresponding desired part of RIR. 

\subsection{Single-channel Dereverberation Neural Networks}
\label{sec:network}
First, we analyze how to model reverberation in the time-frequency domain, based on which a  dereverberation neural network can be properly designed/selected/analyzed. Applying STFT to Eq.~(\ref{eq1}), ignoring the additive noise term and using the narrow-band assumption, we have 
$ Y(k,p)\approx S(k,p)A(k)$,
where \(p\in[1,P]\) and \(k\in[0,K-1]\) denote the frame and frequency indices, $Y(k,p)$ and $S(k,p)$ are the STFT coefficients of $y(n)$ and $x(n)$, $A(k)$ is the Fourier transform of $a(n)$. This assumption is only valid when RIR is short relative to the STFT window, which is obviously not suitable for the dereverberation problem where RIR is normally very long. 

The accurate relation between $Y(k,p)$ and $S(k,p)$ can be represented by the cross-band filters model \cite{LiXiaofei2019Multichannel}:
\begin{equation}
Y(k,p)=\sum_{k'} S(k',p) * A(k, k',p)
  \label{eq9}
\end{equation}
where convolution is applied along $p$. $Y(k,p)$ is a summation over multiple convolutions (between $S(k,p)$ and filter $A(k, k',p)$) across frequencies $k'$. In theory, to make the equation fully valid, $k'$ should take all the frequencies. However, taking the range of $[k-l,k+l]$ (normally $l=4$, determined by the bandwidth of the mainlobe of STFT window) for $k'$ is enough to make the equation sufficiently valid. This means only a few  frequency neighbors of $S(k,p)$ make the main contribution to $Y(k,p)$. Alternatively, we can say that, by taking convolution with the filter $A(k, k',p)$, $S(k,p)$ makes contribution to only a few frequency neighbors of $Y(k,p)$. As for dereverberation, we can apply deconvolution or inverse filtering to $\{Y(k',p)| k'\in[k-l,k+l]\}$ to recover $S(k,p)$, and almost perfect dereverberation performance can be expected as long as the inverse filters are accurate enough. However it is very difficult to estimate the inverse filters in practice. In this work, we employ neural networks to perform dereverbetion.

Although we believe that the proposed RTS learning target is suitable for both single-channel and multichannel dereverberation, this work focuses on the single-channel case as it has already been widely studied in the deep-learning-based dereverberation field. 
In \cite{li2020online}, towards speech denoising, we have proposed a subband LSTM network (SubNet) that inputs the noisy spectra of one frequency and its neighbouring frequencies, and outputs/predicts the clean speech spectra of this frequency. All the frequencies shares the same network. Based on the theoretical analysis above, it seems that this subband network is exactly suitable for speech dereverberation by mimicking the inverse filtering process. 
In \cite{hao2020fullsubnet}, also towards speech denoising, SubNet is improved by combining a full-band network to further exploit the full-band spectral pattern of speech, as the enhanced speech should have a correct spectral pattern across all frequencies. This network, called FullSubNet, includes full-band LSTMs followed by subband LSTMs. We believe that the full-band spectral pattern of speech is also important for dereverberation. For dereverberation, FullSubNet can be seen as a combination of speech spectral regression and subband inverse fitering. 

For comparison, the network of temporal convolutional network with self-attention (TCN-SA) proposed in \cite{Zhao2020Monaural} is also tested, which was recently developed especially for dereverberation and achieved advanced performance. TCN-SA uses TCN to perform full-band nonlinear mapping from reverberant speech to clean speech, with a self-attention pre-processing moudle. 


Overall, we experimentally study to use our previous speech denoising SubNet and FullSubNet for speech derevereberation. More imptortantly, we evaluate the feasibility of the proposed RTS target with various speech dereverberation networks. In this work, the original SubNet and FullSubNet are slighly modified as: (i) the unidirectional LSTM are changed to bi-directional LSTM (BLSTM) for offline dereverberation; (ii) according to TCN-SA, the network input and target are changed to the cubic root of the magnitude of reverberant speech and desired speech, respectively. 


\section{Experiments}  

\subsection{Experimental Setup}  

The REVERB challenge dataset \cite{Kinoshita2016A} is used. Clean speech signals for this dataset come from the WSJCAM0 and MC-WSJ-AV corpus. The reverberation time \(T_{60}\) for three experimental rooms, i.e. small room, medium room and large room, are 0.25 s, 0.5 s and 0.7 s, respectively. The distances between speaker and microphone array are set to either 50 cm (\emph{near}) or 200 cm (\emph{far}). 
Training data are generated by convolving 7861 clean speeches with 24 × 8 measured RIRs of the training set (through online random matching) provided by REVERB challenge. Here, we regard 8-channel RIR as 8 single channel RIRs. Reverberant speech is added air-conditioning noise with a signal-to-noise ratio (SNR) of either 20 dB or 5 dB, to conduct joint derevereberation and denoising. The one-channel SimData of the test set provided by REVERB challenge is used for test, with a SNR of 20 dB or 5 dB as well. 

The sampling rate is 16 kHz. We apply STFT using a 512-point Hamming window with a 256-point frame shift. For training, the signal length is set to a constant value of 3s. The test signals with variant length are directly fed into the network for inference. For the subband network, the number of hidden units is set to 256 for each BLSTM direction. For FullSubNet, the number of hidden units of fullband and subband BLSTM layers are set to 384 and 256 for each direction, respectively. Based on some preliminary experiments, the target $T_{60}$ of RTS is set to \textbf{0.15 s}. Code and audio examples of this paper are available on our website.\footnote{https://audio.westlake.edu.cn/Research/rts.htm}  

\subsection{Baselines and Evaluation Metrics}

To evaluate the effectiveness of the proposed RTS target, two other targets are tested: (i) \emph{direct-path} (ii) direct-path plus 50 ms of early reflections, simply referred to as \emph{early}. 

This work evaluates the human perceptual quality of dereverberated speech. Evaluation metrics include: (i) Perceptual evaluation of speech quality (PESQ) \cite{Rix2001Perceptual}; (ii) Short-Time Objective Intelligibility (STOI) \cite{Taal2011stoi}, an objective evaluation metric of speech intelligibility. (iii) Mean squared error (MSE) between the predicted and target magnitude spectra, which reflects how well the network can solve the given problem. These three metrics are intrusive, and they take the respective target signal as the reference signal for different training targets.
(iv) DNSMOS P.835 \cite{Reddy2022mos}: a neural-network-based non-intrusive speech perceptual quality metric, based on the ITU-T Rec. P.835 subjective evaluation framework, including three ratings: speech quality (SIG), background noise quality (BAK), and the overall audio quality (OVRL). DNSMOS was originally developed for evaluating speech denoising performance. It cannot well measure one important aspect of dereverberation, namely the amount of remained reverberation.

\begin{table*}[t]
\caption{Dereverberation performance.}
\label{tab:target}
\renewcommand\arraystretch{1.2}
\tabcolsep0.05in
\scriptsize
\begin{tabular}{ccccccccccccccc}
\toprule
\textbf{}                   & \textbf{}      & \textbf{}                      & \multicolumn{2}{c}{\textbf{PESQ}$\uparrow$}   & \multicolumn{2}{c}{\textbf{STOI} (\%) $\uparrow$}   & \multicolumn{2}{c}{\textbf{MSE} ($10^{-3}$) $\downarrow$}                                            & \multicolumn{2}{c}{\textbf{MOS-SIG}$\uparrow$} & \multicolumn{2}{c}{\textbf{MOS-BAK}$\uparrow$} & \multicolumn{2}{c}{\textbf{MOS-OVRL}$\uparrow$} \\ 
\textbf{SNR} & \textbf{target}     &  \textbf{network}   & \textbf{unpro.} & \textbf{enhanced} & \textbf{unpro.} & \textbf{enhanced} & \multicolumn{1}{c}{\textbf{unpro.}} & \multicolumn{1}{c}{\textbf{enhanced}} & \textbf{unpro.}  & \textbf{enhanced} & \textbf{unpro.}  & \textbf{enhanced} & \textbf{unpro.}  & \textbf{enhanced} \\ \hline
\multirow{9}{*} {20 dB}   & \multirow{3}{*} {$\emph{direct-path}$}   & \textbf{TCN+SA}  \cite{Zhao2020Monaural} & 1.48            & 2.62              & 85.6            & 93.5              & 9.62                           & 1.67                              & 3.61             & 3.53              & 3.87             & 4.32              & 3.27             & 3.43              \\
 &           & \textbf{SubNet} \cite{li2020online}             &  1.48            & 2.68              & 85.6            & 94.3              & 9.62                            & 1.05                              & 3.61             & 3.54     & 3.87             & 4.34     & 3.27             & 3.45     \\
  &      & \textbf{FullSubNet} \cite{hao2020fullsubnet}        &  1.48            & 2.89     & 85.6            & 94.9     & 9.62                           & 1.03                     & 3.61             & 3.53              & 3.87             & 4.33              & 3.27             & 3.45     \\ \cline{2-15}
 & \multirow{3}{*} {$\emph{early}$}  & \textbf{TCN+SA} \cite{Zhao2020Monaural} &  1.81            & 2.89              & 93.8            & 96.2              & 2.61                            & 1.27                              & 3.61             & 3.68              & 3.87             & 4.34              & 3.27             & 3.56              \\
&&\textbf{SubNet} \cite{li2020online}             &  1.81            & 3.07              & 93.8            & 96.9              & 2.61                          & 0.94                           & 3.61             & 3.72              & 3.87             & 4.36              & 3.27             & 3.60               \\
&&\textbf{FullSubNet} \cite{hao2020fullsubnet}        &  1.81            & 3.22     & 93.8            & 97.1     & 2.61                            & 0.94                     & 3.61             & 3.73     & 3.87             & 4.37     & 3.27             & 3.63     \\ \cline{2-15}
& \multirow{3}{*} {RTS}  & \textbf{TCN+SA} \cite{Zhao2020Monaural} &  1.85            & 3.05              & 93.1            & 96.5              & 3.84                            & 0.93                             & 3.61             & 3.63              & 3.87             & 4.36              & 3.27             & 3.56              \\
&&\textbf{SubNet} \cite{li2020online}             &  1.85            & 3.23              & 93.1            & 97.4              & 3.84                           & 0.43                           & 3.61             & 3.70               & 3.87             & 4.39     & 3.27             & 3.63              \\
&&\textbf{FullSubNet} \cite{hao2020fullsubnet}        &  1.85            & 3.35     & 93.1            & 97.7     & 3.84                           & 0.41                     & 3.61             & 3.72     & 3.87             & 4.38              & 3.27             & 3.64     \\ \hline
\multirow{9}{*} {5 dB}   & \multirow{3}{*} {$\emph{direct-path}$} & \textbf{TCN+SA} \cite{Zhao2020Monaural} &  1.23            & 2.33              & 81.2            & 91.4              & 12.4                           & 1.55                             & 3.50              & 3.46              & 3.57             & 4.26              & 3.12             & 3.32              \\
&&\textbf{SubNet} \cite{li2020online}             & 1.23            & 2.35              & 81.2            & 91.6              & 12.4                            & 1.09                     & 3.50              & 3.47     & 3.57             & 4.30      & 3.12             & 3.35     \\
&&\textbf{FullSubNet} \cite{hao2020fullsubnet}        &  1.23            & 2.50      & 81.2            & 92.7     & 12.4                            & 1.22                              & 3.50              & 3.43              & 3.57             & 4.27              & 3.12             & 3.33              \\ \cline{2-15}
 & \multirow{3}{*} {$\emph{early}$} & \textbf{TCN+SA} \cite{Zhao2020Monaural} &  1.34            & 2.41              & 88.2            & 93.6              & 5.33                            & 1.47                              & 3.50              & 3.61              & 3.57             & 4.25              & 3.12             & 3.43              \\
&&\textbf{SubNet} \cite{li2020online}             &  1.34            & 2.51              & 88.2            & 94.1              & 5.33                            & 1.08                              & 3.50              & 3.61              & 3.57             & 4.33     & 3.12             & 3.48              \\
&&\textbf{FullSubNet} \cite{hao2020fullsubnet}        &  1.34            & 2.68     & 88.2            & 94.8     & 5.33                           & 1.01                     & 3.50              & 3.62     & 3.57             & 4.29              & 3.12             & 3.48     \\ \cline{2-15}
&\multirow{3}{*} {RTS} &\textbf{TCN+SA} \cite{Zhao2020Monaural} &  1.35            & 2.46              & 86.8            & 94.0                & 6.62                            & 0.93                              & 3.50              & 3.59             & 3.57             & 4.30               & 3.12             & 3.45              \\
&&\textbf{SubNet} \cite{li2020online}             & 1.35            & 2.55              & 86.8            & 94.4              & 6.62                            & 0.58                     & 3.50              & 3.56               & 3.57             & 4.34     & 3.12             & 3.47              \\
&&\textbf{FullSubNet} \cite{hao2020fullsubnet}        &  1.35            & 2.73     & 86.8            & 95.3     & 6.62                            & 0.77                             & 3.50              & 3.56     & 3.57             & 4.33              & 3.12             & 3.48     \\ \bottomrule
\end{tabular}
 \vspace{-0.3cm}
\end{table*}

\subsection{Dereverberation Results}

Table~\ref{tab:target} shows the dereverberation performance.  We have the following important findings:

\begin{itemize}[leftmargin=*]
    \item SubNet and FullSubNet outperform TCN+SA for almost all the experimental conditions and evaluation metrics. This verifies that the subband network is indeed suitable for dereverberation, as the convolution between source speech and RIR can be decomposed/modeled within subband, and the subband network can perform kind of deconvolution. Compared to SubNet, FullSubNet further improves the PESQ, STOI scores, although MSE is not reduced. This indicates that although not improve the frequency-wise prediction, FullSubNet improves the spectral correlation between different frequencies, to make the enhanced speech have better fullband spectral pattern and perceptual quality. The superority of FullSubNet is clearly audible when listening to the signals. 
    \item The DNSMOS scores can be largely improved when relaxing the strict \emph{direct-path} target to \emph{early} or the proposed RTS. In specific, mainly the MOS-SIG scores are improved. The high difficulty of predicting direct-path signal leads to a larger speech distortion and thus a worse speech quality. The larger speech distortion of \emph{direct-path} is audible when listening to the signals, especially when SNR is low, i.e. 5 dB.
    \item The \emph{early} target has a smaller MSE than RTS for the unprocessed signal, namely \emph{early} needs to suppress less reverberation than RTS. However, RTS achieves much smaller MSE for the enhanced signal. This means the derevereberation problem can be solved better by using the RTS target, since it is easier for the network to learn a mapping function when the target is also exponentially decaying. As a result, RTS acheives better PESQ and STOI scores than \emph{early}. Moreover, the enhanced signal with exponentially decaying reverberation sounds more natural than the one with suddenly truncated reverberation, in the sense that human never perceive this truncated type of reverberation in real life. Compared to RTS, more and unnatural remaining reverberation can be perceived for \emph{early} when listening to the signals.
    \item The comparisons between different training targets described above are valid for all the three networks. This verifies that the proposed RTS target is suitable for various networks. 
\end{itemize}

\subsection{Analysis of Energy Decay Curve}

To further evaluate the effect of the proposed training target, we analyze the remaining RIR of the enhanced speech \(\tilde x(n)\), which is approximately identified as 
\begin{equation}
\tilde a(n)=\text{Re}\{\operatorname{IDFT}\left[\frac{\operatorname{DFT}(\tilde{x}(n))}{\operatorname{DFT}(s(n))}\right]\}
\label{eq:tan}
\end{equation}
where $\operatorname{DFT}$ and $\operatorname{IDFT}$ denote discrete Fourier transform and inverse DFT, respectively. Then the Energy Decay Curve (EDC) can be obtained based on the Schroeder integral  \cite{Heinrich2000room}: $E D C(n)=\sum_{n'=n}^{N} (\tilde a(n'))^{2} $.

\begin{figure}[t]
  \centering
  \includegraphics[width=0.8\linewidth]{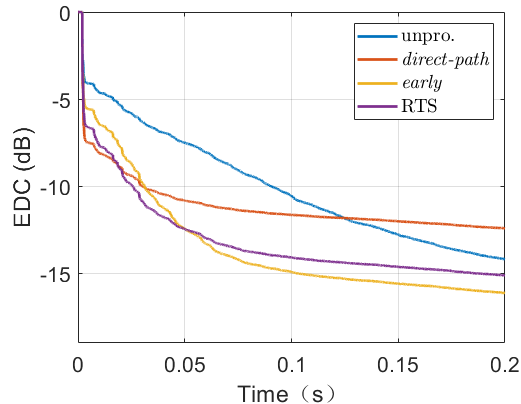}
  \caption{Energy Decay Curves. }
  \label{fig:edc}
  \vspace{-0.3cm}
\end{figure}

Fig.~\ref{fig:edc} shows the EDCs of one example utterance for the unprocessed and enhanced signals with different training targets. The original \(T_{60}\) is 0.7 s, and the distance between speaker and microphone is 2 m, for which case reverberation is very heavy, the background noise is 20 dB. Compared to the unprocessed signal, a huge amount of reverberation have been removed for all enhanced signals. The EDC of enhanced signals decreases rapidly at the early stage,
then the decreasing rate becomes smaller, resulting in a long tail. Note that the long tail is much higher than the  target EDC, as the target EDC linearly decreases along the time axis. The long tail includes both remaining late reverberation and prediction errors. By listening to the enhanced signals, the amount of remaining early reflections can be correctly reflected by EDCs, while the long tail of EDCs are more related to the amount of signal distortion and unnaturalness.
Compared to \emph{early}, the proposed targets has a lower EDC at the early stage and a comparable EDC tail. Compared to \emph{direct-path}, the proposed targets has a slighly higher EDC at the very early stage (earlier than 20 ms), and a much lower EDC tail. Overall, from the perspective of suppressing reverberation and reducing signal distortion, the proposed target is a better choice than \emph{direct-path} and \emph{early}.

\section{Conclusion}

In this paper, we have proposed a reverberation time shortening (RTS) target for speech dereverberation. RTS suppresses reverberation and meanwhile maintains the exponential decay property of reverberation, which improves the sound naturalness and also eases the network training. By adapting the advanced FullSubNet to speech dereverberation, together with the RTS target, outstanding single-channel dereverberation performance has been achieved. 

\footnotesize

\bibliographystyle{IEEEbib}
\bibliography{refs}



\end{document}